\documentclass[
    reprint, 
    amsmath,amssymb,
    aps, pra,
    longbibliography
]{revtex4-2}

\usepackage[english]{babel}
\usepackage{graphicx}
\usepackage{dcolumn}
\usepackage{bm}
\usepackage{hyperref}
\usepackage{xcolor}
\usepackage{float}

\newcommand{\bra}[1]{ \mbox{$\langle\, #1 \,|$}}
\newcommand{\ket}[1]{ \mbox{$|\, #1\,\rangle$}}

\begin{document}

\title{Extracting RABBITT-like phase information from time-dependent transient absorption spectra}                       
\author{J.~Jakob$^{1*}$}
\author{C.~Bauer$^{1}$}
\author{M.-J. Ilhan~$^{1}$}
\author{D.~Bharti$^{2}$}
\author{C.~Ott$^{2}$}
\author{T.~Pfeifer$^{2}$} 
\author{K.~Bartschat$^{3}$} 
\author{A.~Harth$^{1}$}
\email{anne.harth@hs-aalen.de, julian.jakob@hs-aalen.de}

\affiliation{$^1$Center for Optical Technologies, Aalen School of Applied Photonics, Aalen University, D-73430 Aalen, Germany}
\affiliation{$^2$Max-Planck-Institute for Nuclear Physics, D-69117 Heidelberg, Germany}
\affiliation{$^3$Department of Physics and Astronomy, Drake University, Des Moines, IA 50311, USA}

\date{\today} 

\begin{abstract}  
We explore how the spectral phase of atto\-second pulse trains influences the optical cross section in transient absorption (TA) spectroscopy. 
The interaction of extreme ultraviolet (XUV) and time-delayed near-infrared (NIR) fields with an atomic or molecular system governs the dynamics. 
As already shown in \hbox{RABBITT} experiments, the spectral phase of the XUV pulses can be extracted from the photo\-ionization spectrum as a function of the time delay. 
Similarly, this XUV phase imprints itself on delay-dependent optical cross-section oscillations. 
With a perturbative analytical approach and by simulating the quantum dynamics both in a few-level model and via solving the time-dependent Schrödinger equation for atomic hydrogen, we reveal the similarity  between the spectral phase in \hbox{RABBITT} and TA spectroscopy.
\end{abstract}
 
\maketitle        

\section{Introduction}
Atto\-second pulse trains (APTs), which result from the generation of high-order harmonics~\cite{Lewenstein1994,Paul2001,LopezMartens2005}, have become powerful tools for probing ultrafast phenomena in atoms and molecules. 
The original purpose of the Reconstruction of Atto\-second Beating by Interference of Two-Photon Transitions (\hbox{RABBITT}) technique was to utilize extreme ultra\-violet (XUV) and near-infrared (NIR) fields to obtain information about the spectral phase $\Delta \varphi_{\mathrm{XUV}}$ of the atto\-second pulses in the train~\cite{Hentschel_2001}. 

In \hbox{RABBITT}, the APT ionizes the target, leading to the formation of main bands (MBs) in the photo\-electron signal at distinct energies~$E_i$.
In the presence of a time-delayed NIR field, signals appearing in between the MBs, referred to as sidebands (SBs), oscillate at twice the NIR frequency.
This effect arises from the interference of electron wave packets in the continuum ~\cite{Johnsson2007,Dahlstroem2017}. The observed oscillation along the delay contains a phase term $\Delta \phi(E)$ with two distinct contributions~\cite{Kluender2011}:
\begin{equation}
    \Delta \phi(E) = \Delta \varphi_{\mathrm{XUV}} + \Delta\phi_{\mathrm{atom}}(E). 
    \label{eq: 3_phases}
\end{equation}
The first part represents the originally desired spectral dispersion of the XUV pulses, while the second term accounts for an additional phase shift due to interactions involving bound-continuum and continuum-continuum couplings.  The latter have gained increasing attention in recent studies~\cite{Dahlstroem2013,Divya2021,Harth2019,Fuchs2020}.

A complementary method for investigating electron dynamics and ionization processes in atoms and molecules is transient absorption (TA) spectroscopy~\hbox{\cite{Beck2015,Wu2016}}.
In contrast to the \hbox{RABBITT} method, TA spec\-tros\-copy commonly utilizes isolated atto\-second pulses in traditional pump-probe measurement schemes.
The broadband XUV spectra of these pulses enable the ex\-ploration of light-in\-duced states in TA set\-ups~\hbox{\cite{Chen2012, Chen2013kb, Bell2013,Birk2020}}. These isolated pulses also allow for the examination of coherent electron wave packets and their spectral phases~\cite{Dahlstroem2017a}.

APTs can also be utilized in TA spectroscopy.
Chen {\it et al.}~\cite{Chen2012a} and Liao {\it et al.}~\cite{Liao2015} conducted studies using high-order harmonics and observed oscillations at four times the NIR angular frequency $\omega_\mathrm{F}$ in the absorbed spectrum.
These oscillations were attributed to the high NIR intensity in their setup, which induces multi-photon absorption and emission processes involving more than two NIR photons.
Additionally, the authors identified novel spectral features resulting from extended propagation lengths through the gas.
Similarly, Holler {\it et al.}~\cite{Holler2011} investigated high-order harmonics around the ionization threshold of helium and observed $2\,\omega_\mathrm{F}$ oscillations in the photon yield as a function of time delay at positions corresponding to \hbox{RABBITT}-like MBs.
They proposed that the relative phase shift between the MBs is linked to the XUV dispersion $\Delta\varphi_{\mathrm{XUV}}$ but could not validate this hypothesis with their experimental dataset.

The present paper focuses on demonstrating, in general, that MB oscillations in TA spectroscopy also correspond to $\Delta\varphi_{\mathrm{XUV}}$.
Using the same atomic system and identical APTs in this theoretical work, the XUV phase can be extracted through two different approaches: either by analyzing the photo\-ionization spectrum in \hbox{RABBITT} setups or by evaluating the optical cross section in TA spectroscopy. 

\begin{figure}
    \includegraphics[width=1\linewidth]{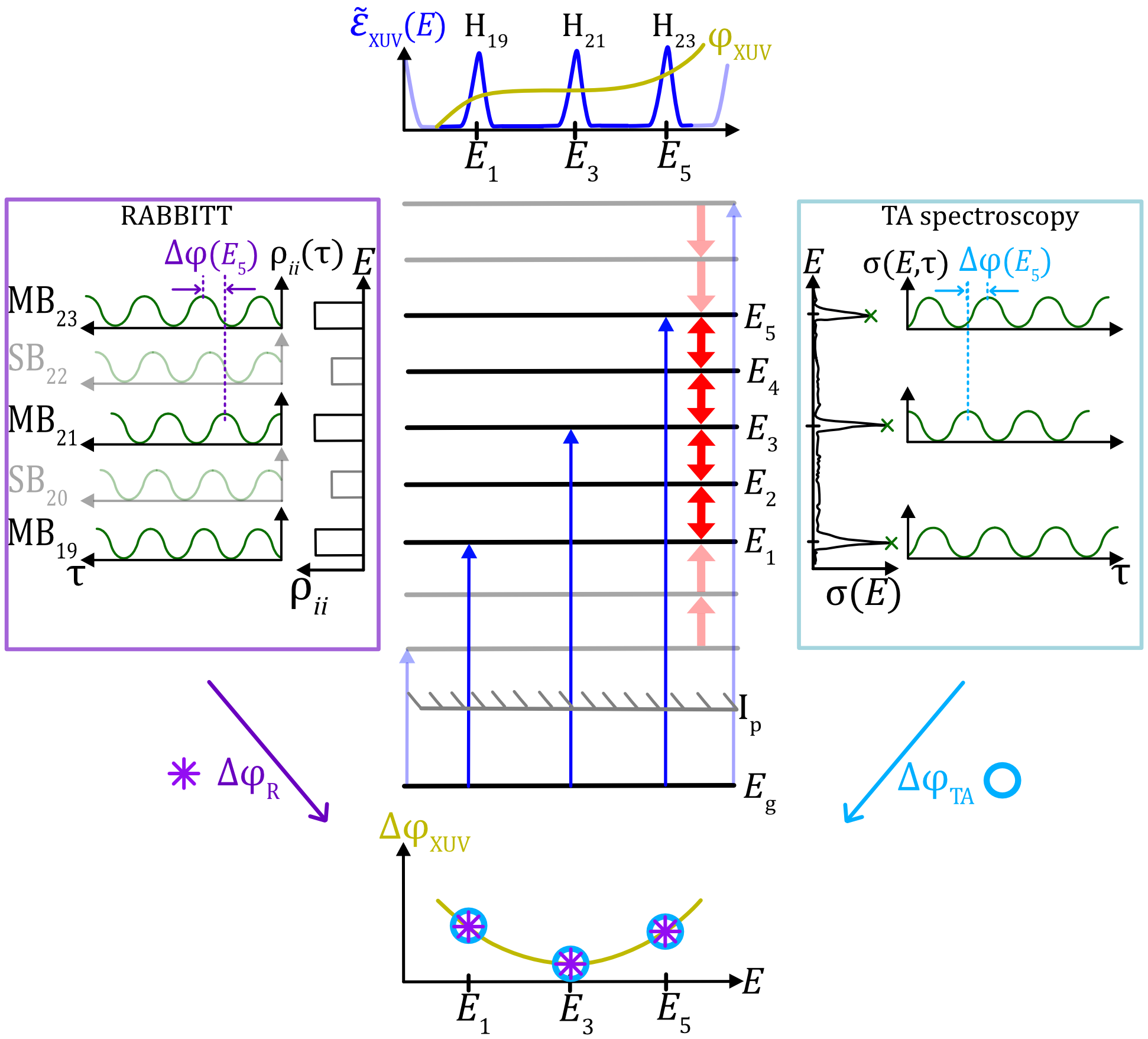}
    \caption{\label{fig: System_skizze1} Schematic representation of the phase extraction via two different approaches. For both \hbox{RABBITT} and TA, the same APT with the same phase $\varphi_\mathrm{XUV}$ and the identical atomic system are depicted. The long thin blue arrows represent the one-photon ionization process driven by the high-order harmonics, while the short red arrows indicate the interactions with the NIR photons. The same phase information can be extracted by either measuring the oscillations of the state population in the MBs (left purple box) or by observing the electric polarization and the resulting cross section (right light blue box) as functions of the time delay.}
\end{figure}

The basic idea is depicted in Fig.~\ref{fig: System_skizze1}.
On top, a sharp XUV spectrum $\tilde{\mathcal{E}}_{\mathrm{XUV}}(E)$ (also referred to as an XUV comb leading to the APT) at energies $E_{1}$, $E_{3}$, and $E_{5}$ corresponding to the high-order harmonics H$_{19}$, H$_{21}$, and H$_{23}$ is shown in blue together with an artificial spectral phase ~$\varphi_{\mathrm{XUV}}$ in yellow.
At the beginning, the target is in its bound ground state~$E_\mathrm{g}$. The interaction with the APT then 
results in the population of continuum states (long thin blue arrows), specifically those with energies $E_1$, $E_3$, $E_5$, which are coupled to each other in the presence of the time-delayed NIR field (indicated by short red arrows). 

Observing the delay-dependent photo\-electron spectrum corresponds to the traditional \hbox{RABBITT} scan (left-side purple box).
However, observing instead the comb-like transmitted XUV spectrum yields a delay-dependent TA spectroscopy map (right-side blue box). 
As shown below, analyzing the oscillations and phase shifts from both data sets separately reveals equivalent relative phase information.

This paper is organized as follows: First, a strongly simplified perturbative analytical description is presented to highlight the equivalence of the retrieved phases.
Next, a numerical simulation based on an artificial few-level system validates the analytical predictions and provides further insights regarding the response of a system to the laser and target parameters.
Finally, the effect of the same pulses is analyzed for the realistic case of atomic hydrogen. Solving the time-dependent Schr\"odinger equation (TDSE) produces results that solidify the previous statements.
We finish the paper with a brief summary.

\section{Analytical approach}
To show the key idea and simplify the mathematics, the total field spectrum $\tilde{\mathcal{E}}_\mathrm{tot}(\omega)$, comprising the probe and pump fields, is expressed as a superposition of delta distributions in the frequency domain:
\begin{align}
    \tilde{\mathcal{E}}_\mathrm{tot}(\omega) =   \tilde{A}_\mathrm{F}\delta(\omega_\mathrm{F} - \omega) {\rm e}^{\rm{i}\omega_\mathrm{F}\tau} +\sum_j \tilde{A}_{j} \delta(\omega_{j}-\omega){\rm e}^{\rm{i}\varphi_{j}}.
\end{align}
The complex XUV field contains multiple frequencies~$\omega_{j}$, each with a spectral XUV phase $\varphi_{\mathrm{XUV}}(\omega_j) \equiv \varphi_j$ and an associated spectral amplitude~$\tilde{A}_{j}$. The complex fundamental NIR field (subscript~``F'' for brevity) \hbox{$\tilde{\mathcal{E}}_\mathrm{F}(\omega)=\tilde{A}_\mathrm{F}\delta(\omega_\mathrm{F} - \omega) {\rm e}^{\rm{i}\omega_\mathrm{F}\tau}$} is assumed to be Fourier-limited, i.e., $\varphi_\mathrm{F}=0$. The NIR field $\mathcal{E}_{\mathrm{F}}(t-\tau)$ is time-delayed by $\tau$. 

In the time domain, the field with its amplitudes $A_\mathrm{F}$ and $A_{j}$ reads
\begin{align}
    \mathcal{E}_\mathrm{tot}(t) =   \underbrace{A_\mathrm{F} {\rm e}^{\rm{i}\omega_\mathrm{F}(\emph{t}-\tau)}}_{:=\mathcal{E}_\mathrm{F}(\emph{t}-\tau)} +\sum_j \underbrace{A_{j} {\rm e}^{\rm{i}(\omega_{j}\emph{t} + \varphi_{j})}}_{:= \mathcal{E}_{j}(t)}.
\end{align}

Consider again the sample system illustrated in Fig.~\ref{fig: System_skizze1}. The delay-dependent population amplitude for 
excitation of all states in the system can be determined using perturbation theory.
However, only the continuum state with energy $E_3$ will be discussed in detail below. 
The quantum paths leading to the state with $E_3$ involve one-photon and three-photon transitions. Therefore, using first- and third-order perturbation theory and denoting the ground state by the subscript ``g'', we obtain
\begin{align}
c_3(t) \propto \mu_\mathrm{g3} \; \mathcal{E}_\mathrm{3}(t) &+ \mu_\mathrm{g5}\mu_\mathrm{54}\mu_\mathrm{43}\; \mathcal{E}_\mathrm{5}(t) (\mathcal{E}_\mathrm{F}^*(t-\tau))^2 \nonumber\\
&+ \mu_\mathrm{g1}\mu_\mathrm{12}\mu_\mathrm{23}\; \mathcal{E}_\mathrm{1}(t) (\mathcal{E}_\mathrm{F}(t-\tau))^2.
\label{eq: c_3(t)}
\end{align}

Three distinct one-photon ionization processes can be identified, each originating from the high-order harmonics H$_{19}$, H$_{21}$, and H$_{23}$ to lead to the respective state with energy $E_i$. Next, to end up in the state with~$E_3$, two NIR photons must be absorbed from the state with $E_\mathrm{1}$,  whereas two NIR photons must be emitted from the state with $E_\mathrm{5}$. Note the complex conjugate (indicated by the asterisk in Eq.~(\ref{eq: c_3(t)})) of the fundamental field in the emission path. Also, the couplings between the states with $E_i$ and $E_j$ are weighted with the dipole matrix elements~$\mu_{ij}$.

\subsection{Transient Absorption Spectroscopy}
The electric polarization can be expressed by the trace of the product of the dipole operator $\hat{\mu}$ and the density operator $\rho$: $P(t)\propto tr\left(\hat{\mu} \rho(t) \right)$. 
Continuum-continuum couplings occur at spectral energies within the infrared region. 
However, for isolating the influence of the XUV comb, only the bound-continuum couplings are relevant. 
Then, the transmitted spectrum only contains contributions that couple to the ground state of the system:
\begin{align}
P(t) &\propto \sum_j \mu_{\mathrm{g}j} \, \rho_{\mathrm{g}j}(t)  + c.c. \propto \sum_j P_j(t).
\label{eq_opticalP}  
\end{align}
The dipole matrix elements $\mu_{\mathrm{g}j}$ vanish for all couplings that are not allowed by the selection rules for electric dipole radiation.

To simplify the approach, we now analyze the polarization by considering only one contribution corresponding to a specific energy difference, namely the unique difference from the ground state to~$E_3$. 
In our simplified model, this condition is satisfied for the energy levels of interest. 
In the perturbative case, the amplitude of the ground state can be assumed to not vary significantly in time, i.e., $|c_g(t)|\approx 1$ at all times. This allows us to write
\begin{equation}
    P_3(t) \approx \mu_\mathrm{g3} c_3(t) + c.c.
\end{equation}
Inserting Eq.~(\ref{eq: c_3(t)}) for $c_3(t)$, factoring out the $\mathcal{E}_\mathrm{3}(t)$ term, and noting that $\omega_\mathrm{5/1} = \omega_\mathrm{3} \pm 2\,\omega_\mathrm{F}$, it follows that
\begin{align}
    P_3(\omega) = \Big( \mu_\mathrm{g3}^2     
            &+ \mu_\mathrm{g3}\mu_\mathrm{g5}\mu_\mathrm{54}\mu_\mathrm{43} {\rm e}^{\rm{i}(\varphi_{5} -\varphi_{3})} 
            \tilde{A}_\mathrm{F}^{*2} {\rm e}^{2\,\rm{i}\,\omega_\mathrm{F}\tau}     \nonumber\\
            &+ \mu_\mathrm{g3}\mu_\mathrm{g1}\mu_\mathrm{12}\mu_\mathrm{23} {\rm e}^{\rm{i}(\varphi_{1} -\varphi_{3})}
            \tilde{A}_\mathrm{F}^{2} e^{-2\,\rm{i}\,\omega_\mathrm{F}\tau}  \Big) \nonumber \\
            &\cdot \tilde{A} {\rm e}^{\rm{i}\varphi_{3}} \delta(\omega_3-\omega)
\end{align}
in the frequency domain.
Here we assumed that all XUV amplitudes are the same, i.e., $\tilde{A}_j \equiv \tilde{A}$. The Fourier transform of the complex conjugate can be neglected in this context, as the field $\mathcal{E}_\mathrm{tot}(t)$ contains only positive frequency components. For the sake of simplicity and to focus only on the XUV phase, the dipole moments are all chosen to be the same and real, which implies that the atomic phase $\Delta \phi_\mathrm{atom}=0$. With $E = \hbar\omega$, the optical cross section~\cite{Beck2015} at the energy level $E_3$ is then given by 
\begin{align}
    \sigma(E_\mathrm{3}, \tau) &\propto E_\mathrm{3} \, \Im\left( \frac{P_3(E_\mathrm{3})}{\mathcal{E}_\mathrm{tot}(E_\mathrm{3})} \right) \nonumber\\
    &\propto E_\mathrm{3} \big[\sin(\varphi_{5} -\varphi_{3} + 2\,\omega_\mathrm{F} \tau )\nonumber \\
    &\;\;\ + \sin(\varphi_{1} -\varphi_{3} -2\,\omega_\mathrm{F} \tau) \big],
\end{align}
where $\Im(z)$ denotes the imaginary part of the complex quantity~$z$.

Using trigonometric addition theorems, the characteristic $2\,\omega_\mathrm{F}$ oscillation in $\tau$, along with a phase shift introduced by the XUV pulse, becomes evident in the cross section
\begin{align}
    \sigma(E_\mathrm{3}, \tau) \propto E_\mathrm{3}
    \cos\bigg(2\,\omega_\mathrm{F}\tau - \underbrace{\frac{1}{2} (\varphi_5-\varphi_1}_{:=\Delta\varphi(\omega_\mathrm{3})}) \bigg) .
    \label{eq: opt density phase}
\end{align}
With this scheme, the oscillation phases in $\tau$ can also be determined for other energy levels corresponding to higher-order harmonic energies beyond the example of $E_3$. 
This behavior is sketched in the light blue box in Fig.~\ref{fig: System_skizze1}, where delay-dependent oscillations in the cross section at energies $E_1$ and $E_5$ with phases $\Delta \varphi(\omega_1)$ and $\Delta \varphi(\omega_5)$ appear as well.

\subsection{RABBITT}
In contrast to a TA setup, delay-dependent signal oscillations in \hbox{RABBITT} experiments are observed by measuring the photo\-electron spectrum. This spectrum directly reflects the population of the states involved. For comparison, we now calculate the MB at $E_3$ explicitly.
Abbreviating each term from Eq.~(\ref{eq: c_3(t)}) with $a,b,c$, respectively, the population probability can be expressed as
\begin{align}
    \rho_{33}(t) &= |c_3(t)|^2 = |a+b+c|^2 \nonumber\\
    &= |a|^2 + |b^2| + |c|^2 + 2\Re(a^*b) + 2\Re(a^*c) + 2\Re(b^*c), 
\end{align}
where $\Re(z)$ denotes the real part of the complex quantity~$z$.

Using once again $\omega_\mathrm{5/1} = \omega_\mathrm{3} \pm 2\,\omega_\mathrm{F}$ and assuming that the dipole moments $\mu_{ij} := \tilde{\mu}$ are all the same and real, the density matrix element at the end of the pulse ($t = T$) is
\begin{align}
    \rho_{33}(t=T,\tau) &\propto B\cos(\varphi_5-\varphi_3 - 2\,\omega_\mathrm{F}\tau) \nonumber \\
    &+ B \cos(\varphi_1-\varphi_3 + 2\,\omega_\mathrm{F}\tau) \nonumber \\
    &+ B^2\cos(\varphi_5-\varphi_1 + 4\omega_\mathrm{F}\tau) \nonumber \\ 
& \approx B\cos\left(2\,\omega_\mathrm{F}\tau - \Delta\varphi(\omega_\mathrm{3}) \right).
    \label{eq: phase RABBBITT}
\end{align}
Here we neglected the last term, since \hbox{$B = |\tilde{\mu}^3 \tilde{A}^2 \tilde{A}_\mathrm{F}^2| \ll 1$} and used again  addition theorems to reveal the characteristic $2\,\omega_\mathrm{F}$ oscillation.

Note that the \hbox{RABBITT} phase in Eq.~(\ref{eq: phase RABBBITT}) is  the same phase as in the cross section belonging to the TA setup described by Eq.~(\ref{eq: opt density phase}). 
The above calculation can be repeated for all MBs. The approach is sketched in the purple box in Fig.~\ref{fig: System_skizze1}, where delay-dependent oscillations in the population again appear at energies $E_1$ and $E_5$ with phases $\Delta \varphi(\omega_1)$ and $\Delta \varphi(\omega_5)$, respectively.

\section{Few-Level Model Simulation}
In this section, the principal concepts of the method are demonstrated by simulating a quantum-mechanical few-level model in an artificial system. The simplified model contains a limited number of discrete energy levels, encompassing the ground state and a selection of continuum states.
Describing the continuum by a reduced number of states with energies $E_i$ is adequate to show the key idea, as the perturbative interactions with the comb-structured XUV spectrum can be approximated using only a few effective states~\cite{Harth2019}. 

In order to investigate the population probability \hbox{$\rho_{ii}(t) = |c_i(t)|^2$} of each state and the electric polarization~$P(t)$, it is necessary to calculate the complex-valued state amplitudes $c_i(t)$ of the superposition wave function
\begin{equation}
    \ket{\psi(t)} = \sum_i c_i(t) \ket{\phi_i}.
\end{equation}
 The index $i$ runs over all states (ground and continuum). The matrix representation of the unperturbed Hamiltonian $H_\mathrm{0}$ is diagonal in the basis $\{\ket{\phi_i}\}$ i.e., $\bra{\phi_i}H_\mathrm{0}\ket{\phi_j} = \delta_{ij}\hbar \omega_{ij}$.
 
 Next, the states are perturbatively excited by a XUV pulse $\mathcal{E}_\mathrm{XUV}(t)$ and then coupled by a time-delayed NIR pulse $\mathcal{E}_\mathrm{F}(t-\tau)$. The total wave function is propagated in time by an Euler split-step algorithm, which solves the TDSE
\begin{equation}
    \mathrm{i}\hbar \frac{\mathrm{d}}{\mathrm{d t}} \ket{\psi(t,\tau)} = \left(\hat{H}_\mathrm{0} + \hat{H}_\mathrm{int}(t-\tau)\right) \ket{\psi(t,\tau)}
\end{equation}
numerically. The interaction Hamiltonian in the dipole approximation is given by $\hat{H}_\mathrm{int}(t-\tau) = \hat{\mu} \mathcal{E}_\mathrm{F}(t-\tau)$, where the operator $\hat{\mu}$ contains all dipole moments between the states, such that $\bra{\phi_i}\hat{H}_\mathrm{int}(t-\tau) \ket{\phi_j} = \mu_{ij} \mathcal{E}_\mathrm{F}(t-\tau)$. All allowed dipole couplings $\mu_{ij}$ are set the same in this simulation.
Since the time propagation allows the tracking of all state amplitudes $c_i(t,\tau)$, it is possible to calculate the density matrix $\rho(t,\tau) = \ket{\psi(t,\tau)} \bra{\psi(t,\tau)}$. Therefore, the electric polarization for different time delays~$\tau$ and,
after the interaction with both pulses is completed at $t=T$, the population $\rho_{ii}(T,\tau)$ of every state~$i$ associated with the \hbox{RABBITT} setup is recorded.
\begin{figure}
    \centering
    \includegraphics[width = 1\linewidth]{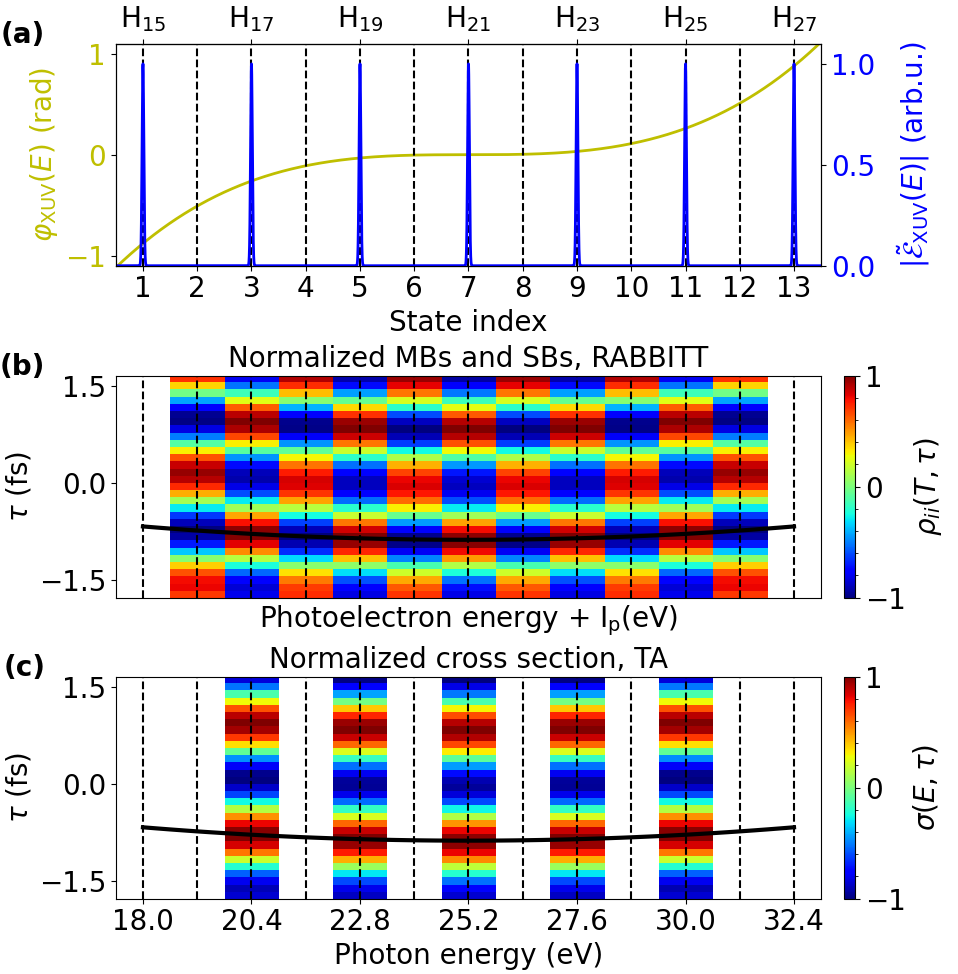}
    \caption{\label{fig: Method1} 
    Initial system and results of the few-level simulation. (a)~Harmonics of the ATP plotted with an applied third-order dispersion. (b)~Population probability of all SBs and MBs normalized and plotted along the time delay~$\tau$. (c)~Optical cross section, also normalized, plotted for the different energies corresponding to the harmonics of the APT. The black line is included to illustrate the parabolic form of the phases. The simulation parameters are: \hbox{$E_\mathrm{F} = 1.2$\,eV} (1030\,nm), central harmonic \hbox{$E_\mathrm{H_{21}}=25.2$\,eV}, XUV dispersion ${\mathrm{TOD} = 1.41 \cdot 10^{-2} \, \mathrm{rad}}/ \mathrm{(eV)}^3$, modulus of all coupling elements \hbox{$|\mu_{ij}| = 0.37$\,a.u.}, APT intensity $10^{-8}$\,a.u. \hbox{($3.51\cdot10^{8}\,\rm W/cm^2$)}, NIR intensity $10^{-6}$\,a.u. \hbox{($3.51\cdot10^{10}\,\rm W/cm^2$)}.}
\end{figure}
\begin{figure*}[ht]
    \centering
    {\includegraphics[width=\textwidth]{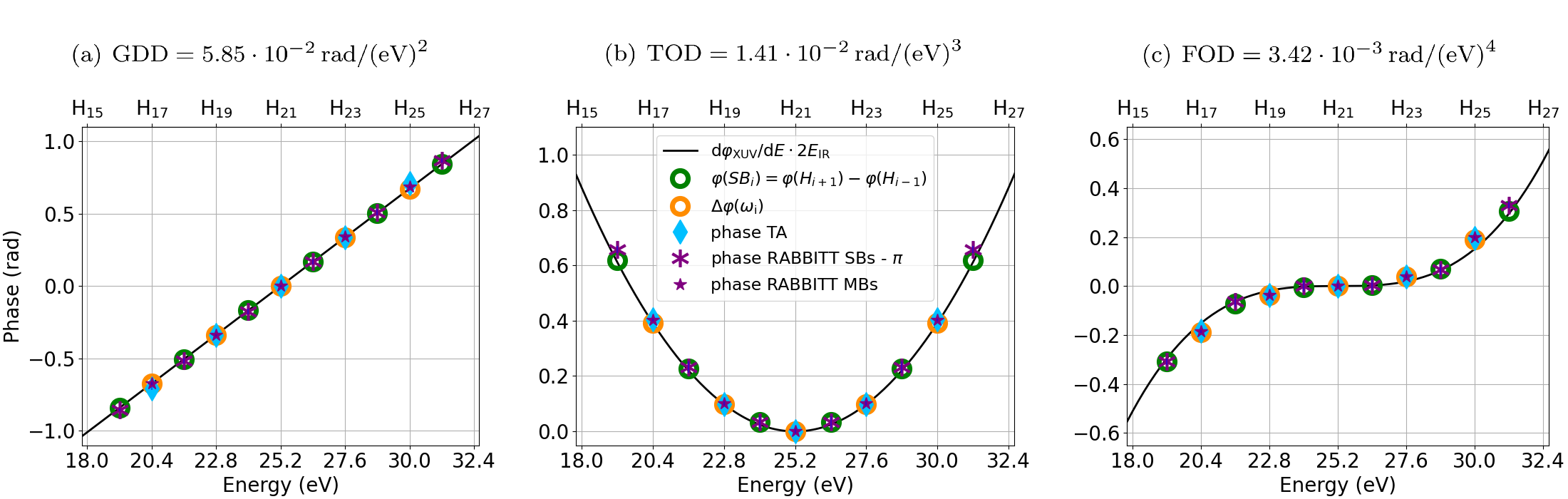}}
    \caption{Phases plotted as a function of the MB and SB energies, obtained by fitting a cosine function to the population and cross-section oscillations for all excited states in the few-level system. The dispersion of the XUV pulse is characterized by GDD, TOD, or FOD. The black line represents the approximation from Eq.~(\ref{eq: diff_approx}), while the orange circles correspond to the theoretical phases for the MB energies from Eqs.~(\ref{eq: opt density phase}) and~(\ref{eq: phase RABBBITT}). The theoretical phases for the SBs are shown with green circles, and the results for the cross section of the simulation as light blue diamonds. The purple stars indicate the \hbox{RABBITT} SB and MB simulation results.}
    \label{fig:few_level_results}
\end{figure*}
Specifically, we set up a system with 14 energy levels, as shown in Fig.~\ref{fig: Method1}(a). The excited states are indicated by the black dotted lines. With the ground state located at $E_0 = 0$\,eV, seven states at the peak energies of the XUV harmonics H$_{15}$ to H$_{27}$ and six levels at the SB positions in between the harmonics compose the system. The spectral amplitudes of the XUV field remain identical, and a third-order dispersion (TOD) is applied as the spectral XUV phase $\varphi_\mathrm{XUV}$. The photons in the NIR pulse have an energy of $1.2\,$eV, and the time delay is chosen to span one full period of the pulse. The delay is symmetrically centered at zero, coinciding with the peak of the XUV pulse envelope. 

The results of the few-level simulation are illustrated in Figs.~\ref{fig: Method1}(b,c). In panel~(b), the normalized oscillations of the photo\-electron spectrum at each state along the $\tau$-direction distinctly demonstrate the expected $2\,\omega_\mathrm{F}$ frequency. Additionally, the phase difference described by Eq.~(\ref{eq: phase RABBBITT}) is already apparent and emphasized by the black parabola in the plot.
Both the MBs at the harmonic energies and the SBs exhibit this frequency dependence with the predicted phase, requiring only a $\pi$ phase shift for alignment. This shift is attributed to the fact that one more photon is necessary to couple the MBs with the SBs.
The normalization is implemented to highlight the phase characteristics rather than the numerical values of the amplitude.

In panel~(c), the optical cross section is normalized along $\tau$ as well. Twice the NIR frequency is once again evident in every energy slice, corresponding to the period of $1.72\,$fs visible in the plot. Only the MB signals are seen because the parity of the ground state prevents contributions to the polarization at the SB energies.

Note that phases from the outer MB cannot be compared with the theoretical values, as only a single coupling with another MB is present at these states.  However, two couplings are required for the correct physical description. The same applies to the outer states in the TA approach. 

It is now possible to extract the resulting phases by performing a cosine fit to the oscillations shown in Fig.~\ref{fig: Method1}. Figure~\ref{fig:few_level_results} exhibits these TA and \hbox{RABBITT} phases for three different applied XUV spectral phases. The theoretical phases from Eq.~(\ref{eq: opt density phase}) or Eq.~(\ref{eq: phase RABBBITT}) can be approximated as
\begin{align}
    \frac{\Delta\varphi_\mathrm{XUV}}{\Delta E} = \frac{2\Delta \varphi(E_j)}{4E_\mathrm{F}} &\approx \frac{\partial\varphi_\mathrm{XUV}}{\partial E} \nonumber \\
    &\Leftrightarrow \Delta \varphi(E_j) \approx \frac{\partial\varphi_\mathrm{XUV}}{\partial E} \cdot 2E_\mathrm{F}.
    \label{eq: diff_approx}
\end{align}
This approximation is shown as the solid black line in Fig.~\ref{fig:few_level_results}, with panels~(a), (b), and~(c) representing a group-delay dispersion $\mathrm{GDD}= 5.85 \cdot 10^{-2} \,\mathrm{rad}/(\mathrm{eV})^2$, a third-order dispersion $\mathrm{TOD} = 1.41 \cdot 10^{-2} \, \mathrm{rad}/(\mathrm{eV})^3$, and a fourth-order dispersion $\mathrm{FOD} = 3.42\cdot 10^{-3}\,\mathrm{rad}/(\mathrm{eV})^4$, respectively.
In each plot, the SB and MB phases from the \hbox{RABBITT} simulation are marked with purple stars, while the phases of the optical cross section are represented by light blue diamonds. All the evaluated oscillations align very well with the expected theoretical values without approximation (marked by green circles for the SBs and orange circles for the MBs).  The error bars from the fit are negligible, thus verifying that the approximation is accurate.

\section{TDSE Simulation}
In this section, a more realistic scenario is examined: an \emph{ab-initio} solution of the TDSE for atomic hydrogen. In this case, the population probabilities at any time during the interaction with the pulses can be accessed, and the polarization (TA) and ionization probability (\hbox{RABBITT}) at the excitation energies of the XUV comb can be investigated for any time delay~$\tau$. The problem is addressed in full dimensionality within the non\-relativistic Schr\"odinger picture, which should be an excellent representation for the situation of interest. 
The numerical solution was carried out on a discretized space-time grid~\cite{Douguet2016}, with recent significant improvements in the numerical aspects~\cite{Douguet2024}. 

Specifically, the calculation was performed with the following parameters: We set up the fundamental IR pulse with a central wavelength of 1,030\,nm (1.2\,eV), a sin$^2$ envelope of 20~cycles ramp-up and 20~cycles ramp-down, and a peak intensity of $10^{11}\,\rm W/cm^2$. In addition, we defined the high-order harmonics $\rm H_{15}, H_{17}, \ldots, H_{27}$ with equal peak intensities of $10^9\,\rm W/cm^2$.  In the time domain, the TOD corresponds to a constant phase that was imposed on the various high-order harmonics symmetrically around $\rm H_{21}$.  Finally, 21 time delays from $-0.5$ to $+0.5$ in steps of 0.05~IR periods were considered.

The time-dependent induced dipole moment, corresponding to the electric polarization $P(t)$ with the electron charge as $-1$ in atomic units, 
\begin{equation}
\langle \,\mu\,\rangle(t) = - \langle \Psi(\bm{r},t)|z|\Psi(\bm{r},t)\rangle,
\end{equation}
was obtained while propagating the initial wave function under the influence of the electric field. The energy-differential ejected-electron emission spectrum was generated by squaring the magnitude of the overlap between the final-state wave function after the interaction and the Coulomb function corresponding to the electron's energy.  Finally, the time-dependent quantities were Fourier-transformed to generate the quantities of interest for the present work.

Since both the optical cross section and the ionization probability can be computed accurately for atomic hydrogen, this system provides an ideal platform for testing the concepts discussed above. 
The XUV phase was initially set to zero, so that the phase contribution $\Delta \phi_\mathrm{atom}$ from the atom  could be extracted to establish a reference system. The analytic formula for this phase is given in Ref.~\cite{Dahlstroem2013,Divya2021}. Subsequently, the same TOD as in the few-level simulation was applied to the XUV pulse.

\begin{figure}[t]
    \centering
    \includegraphics[width = 0.9999\linewidth]{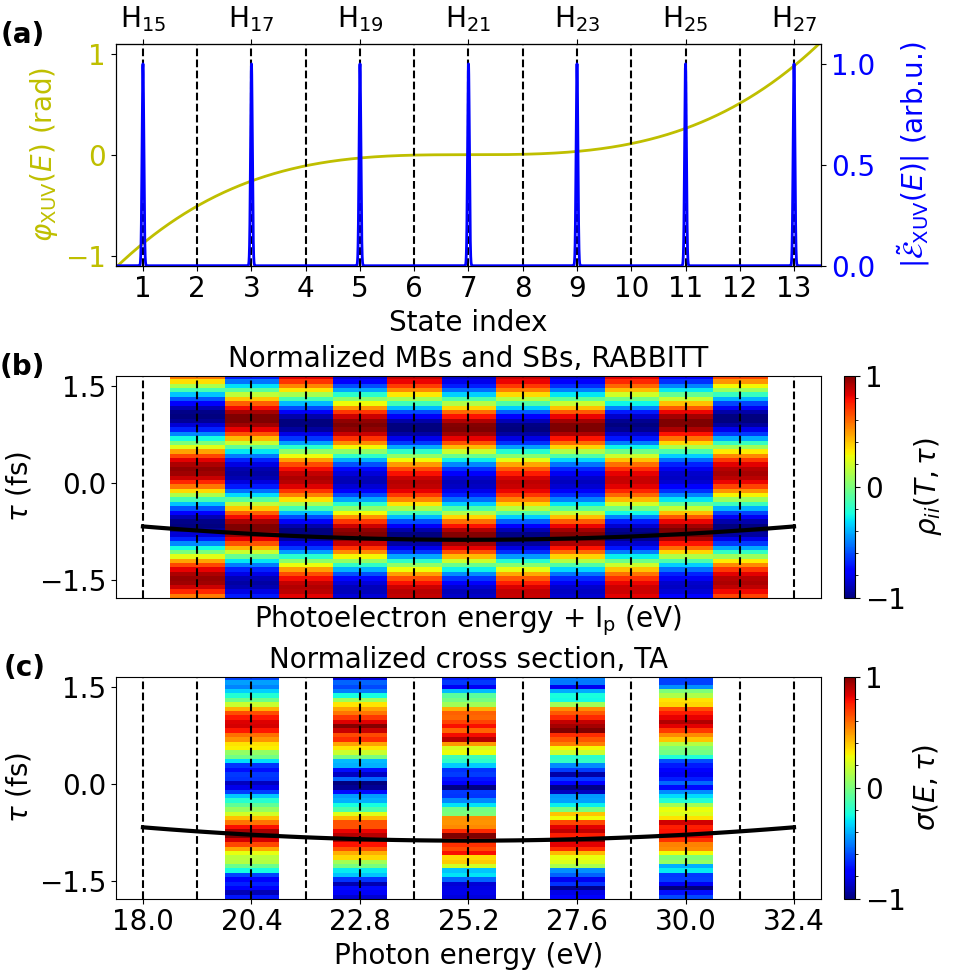}
    \caption{\label{fig: Method2} 
    Initial system and results of the simulation for atomic hydrogen. (a)~Harmonics of the ATP plotted with an applied TOD. (b)~Population probability of all SBs and MBs normalized and plotted along the time delay~$\tau$. (c)~Optical cross section, also normalized, plotted for the different energies corresponding to the harmonics of the APT. The black line is included to illustrate the parabolic form of the phases. 
    The parameters are: $E_\mathrm{F} = 1.2\,$eV (1,030\,nm), central harmonic $E_\mathrm{H_{21}}=25.2\,$eV, XUV dispersion $\mathrm{TOD} = 1.41 \cdot 10^{-2} \mathrm{rad}/\mathrm{(eV)}^3$, APT intensity $2.85 \cdot 10^{-8}$\,a.u. ($10^{9}\,\rm W/cm^2$), NIR intensity $2.85 \cdot 10 ^{-6}$\,a.u. ($10^{11}\,\rm W/cm^2$).
    }
\end{figure}
\begin{figure}[t]
    \centering
    \includegraphics[width = 0.80\linewidth]{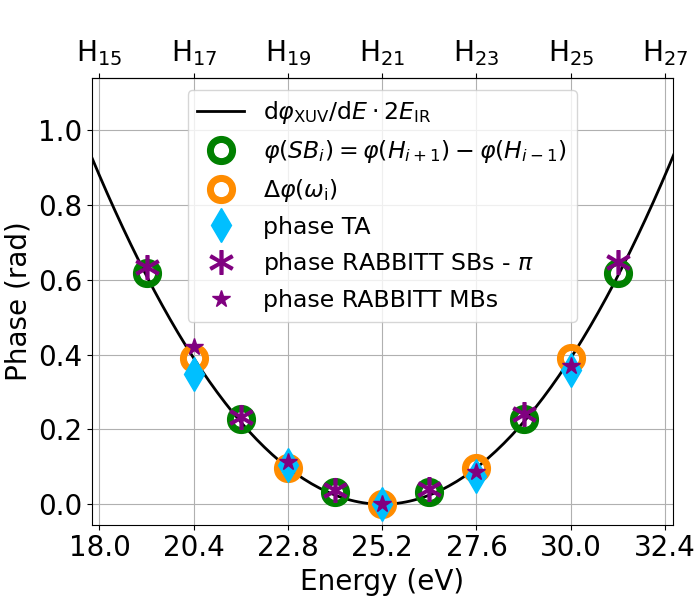}
    \caption{\label{fig: results_method2} Phases plotted as a function of the MB and SB energies, derived by fitting a cosine function to the population and cross-section oscillations obtained in the TDSE calculation for atomic hydrogen. The black line represents the approximation from Eq.~(\ref{eq: diff_approx}), with the orange circles corresponding to the theoretical phases for the MB energies from Eqs.~(\ref{eq: opt density phase}) and~(\ref{eq: phase RABBBITT}). The theoretical phases for the SBs are shown as green circles, while the simulation results for the cross section are plotted as light blue diamonds. The purple stars denote the \hbox{RABBITT} SB and MB simulation results.}
\end{figure}

The results, similar to those obtained with the few-level method, are shown in Fig.~\ref{fig: Method2}. The influence of the XUV phase and the atomic phase are still superimposed, as is evident from the deviation of the oscillation phases of the energy bands from the black parabola, especially at lower energies.
Figure~\ref{fig: results_method2} presents the phase values, with the reference contributions subtracted from the obtained data.
In both cases, with and without an applied XUV phase, the phase is retrieved by fitting a cosine function to the known $2\,\omega_\mathrm{F}$ oscillations. Using the same theoretical values and parabolic approximation as in the few-level model shown in Fig.~\ref{fig:few_level_results}(b), the phases retrieved from the \hbox{RABBITT} and TA analysis closely match the previously obtained values in the few-level model. The same markers and colors are used to represent the results in the plot.

\section{Summary}
In this paper, we demonstrated that the spectral phase of the APT, previously established as measurable in \hbox{RABBITT} experiments~\cite{Hentschel_2001,Divya2021}, can also be extracted through optical-density or cross-section analysis of the transmitted spectrum in TA spectroscopy. While \hbox{RABBITT} relies on measuring the energy of the photo\-electrons and is thus limited to continuum states, TA spectroscopy could extend this capability to regions near the ionization threshold~\cite{Birk2020} and even to bound states. This enables not only the determination of the spectral XUV phase but also the investigation of bound-bound phase dynamics, thereby offering novel insights into atomic and molecular processes.

The key ideas are based on a simplified few-level model with real and constant dipole matrix elements~$\mu_{ij}$. A more comprehensive and realistic description can be achieved by relaxing this assumption, allowing energy-dependent complex dipole matrix elements and thus incorporating atomic phase contributions, as well as including quasi bound states.
Finally, the predictions were solidified by calculations for the real hydrogen atom, taking into account additional effects due to the XUV and NIR phases, such as continuum-continuum couplings and (atomic) Wigner phase contributions.

\begin{acknowledgments}
This work was supported, in part, by Aalen University of Applied Sciences (J.J., C.B, M-J.I, A.H.) and
the United States National Science Foundation under grants PHY-2110023, PHY-2408484, and the ACCESS super\-computer allocation PHY-090031 (K.B.).
Publication funded by Aalen University of Applied Sciences.
\end{acknowledgments}

\end{document}